\documentclass[9pt, sigconf,screen=true,bookmarks=false]{acmart}

\usepackage{color,xcolor}
\usepackage{epsfig}
\usepackage{graphicx}

\usepackage{adjustbox}
\usepackage{array}
\usepackage{booktabs}
\usepackage{colortbl}
\usepackage{float,wrapfig}
\usepackage{hhline}
\usepackage{multirow}
\usepackage{amsfonts}
\usepackage{mathtools}
\usepackage{bm}
\usepackage{nicefrac}
\usepackage{braket}

\usepackage{changepage}
\usepackage{extramarks}
\usepackage{fancyhdr}
\usepackage{setspace}
\usepackage{soul}
\usepackage{xspace}
\usepackage{textcomp}
\usepackage[subrefformat=parens,labelformat=parens]{subfig}

\usepackage{enumerate}
\usepackage{fancyhdr,graphicx,amsmath}
\usepackage[algo2e, ruled, vlined, noend]{algorithm2e}

\usepackage{optidef}
\usepackage{enumitem}

\usepackage{nccfoots}

\usepackage{fancyhdr}

\newcommand{\ignorethis}[1]{}

\makeatletter
\DeclareRobustCommand\onedot{\futurelet\@let@token\@onedot}
\def\@onedot{\ifx\@let@token.\else.\null\fi\xspace}

\definecolor{MyDarkBlue}{rgb}{0,0.08,1}
\definecolor{MyDarkGreen}{rgb}{0.02,0.6,0.02}
\definecolor{MyDarkRed}{rgb}{0.8,0.02,0.02}
\definecolor{MyDarkOrange}{rgb}{0.40,0.2,0.02}
\definecolor{MyPurple}{RGB}{111,0,255}
\definecolor{MyRed}{rgb}{1.0,0.0,0.0}
\definecolor{MyGold}{rgb}{0.75,0.6,0.12}
\definecolor{MyDarkgray}{rgb}{0.66, 0.66, 0.66}

\newcommand{\name}{QOC\xspace}
\newcommand{\pqc}{PQC\xspace}

\newcommand{\torchquantum}{TorchQuantum}

\newcommand{\qnn}{QNN\xspace}

\newcommand{\E}{\mathbb{E}}

\newcommand{\nisq}{NISQ\xspace}

\usepackage{geometry}
\copyrightyear{2022} 
\acmYear{2022} 
\setcopyright{rightsretained}
\acmConference[DAC '22]{Proceedings of the 59th ACM/IEEE Design Automation Conference (DAC)}{July 10--14, 2022}{San Francisco, CA, USA}
\acmBooktitle{Proceedings of the 59th ACM/IEEE Design Automation Conference (DAC) (DAC '22), July 10--14, 2022, San Francisco, CA, USA}
\acmDOI{10.1145/3489517.3530495}
\acmISBN{978-1-4503-9142-9/22/07}

\begin{document}
\settopmatter{printacmref=false} 



\pagestyle{plain}
\title{QOC: \underline{Q}uantum \underline{O}n-\underline{C}hip Training with \\ Parameter Shift and Gradient Pruning}


\author{$^1$Hanrui Wang$^{*}$, $^2$Zirui Li$^{*}$, $^3$Jiaqi Gu, $^4$Yongshan Ding, $^3$David Z. Pan, $^1$Song Han\\
\small{$^1$Massachusetts Institute of Technology, 
$^2$Rutgers University, 
$^3$University of Taxes at Austin, 
$^4$Yale University\\
\texttt{\url{https://qmlsys.mit.edu}}}
}

  



\begin{abstract}
Parameterized Quantum Circuits (\pqc) are drawing increasing research interest thanks to its potential to achieve quantum advantages on near-term Noisy Intermediate Scale Quantum (\nisq) hardware. In order to achieve \emph{scalable} \pqc learning, the training process needs to be offloaded to real quantum machines instead of using exponential-cost classical simulators.
One common approach to obtain \pqc gradients is \emph{parameter shift} whose cost scales \emph{linearly} with the number of qubits. We present \name, the first experimental demonstration of practical on-chip \pqc training with parameter shift.
Nevertheless, we find that due to the significant quantum errors (noises) on real machines, gradients obtained from na\"ive parameter shift have low fidelity and thus degrading the training accuracy. 
To this end, we further propose \emph{probabilistic gradient pruning} to firstly identify gradients with potentially large errors and then remove them. 
Specifically, small gradients have larger relative errors than large ones, thus having a higher probability to be pruned. 
We perform extensive experiments with the Quantum Neural Network (\qnn) benchmarks on 5 classification tasks using 5 real quantum machines. The results demonstrate that our on-chip training achieves over \textbf{90\%} and \textbf{60\%} accuracy for 2-class and 4-class image classification tasks. The probabilistic gradient pruning brings up to \textbf{7\%} \pqc accuracy improvements over no pruning. Overall, we successfully obtain similar on-chip training accuracy compared with noise-free simulation but have much better training \textit{scalability}. The \name code is available in the \href{https://github.com/mit-han-lab/torchquantum}{\torchquantum} library.

\end{abstract}


\maketitle

\section{Introduction}
\label{sec:intro}

Quantum Computing (QC) has great potential to achieve exponential acceleration over classical computers, which represents a computational paradigm shift in various domains. Parameterized Quantum Circuits (PQC) are circuits containing trainable weights and are promising to achieve quantum advantages in current devices. Among them, Quantum Neural Network (QNN) is one of the popular algorithms for machine learning tasks.

\begin{figure}[t]
    \centering
    \includegraphics[width=0.5\textwidth]{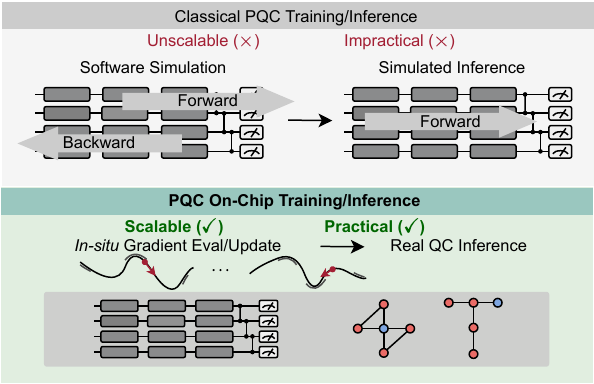}
    \vspace{-15pt}
    \caption{In \name, \pqc training and inference are both performed on \emph{real quantum machines}, making the whole pipeline \textit{scalable and practical}.}
    \label{fig:teaser}
    \vspace{-10pt}
\end{figure}

\begin{figure}
    \centering
    \subfloat[]{\includegraphics[width=0.95\columnwidth]{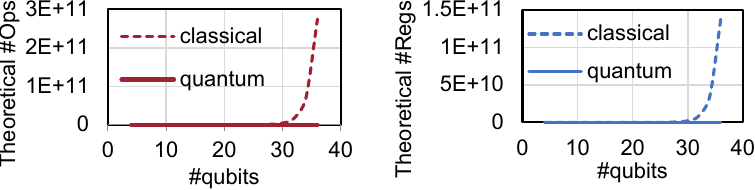}
    \label{fig:theoretical_cost}
    }\\
    \vspace{-10pt}
    \subfloat[]{\includegraphics[width=0.48\columnwidth]{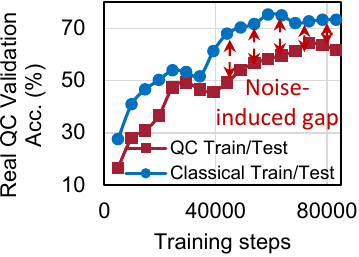}
    \label{fig:classical_quantum_gap}
    }
    \subfloat[]{\includegraphics[width=0.47\columnwidth]{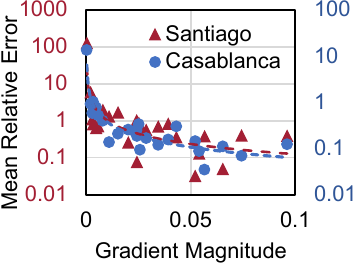}
    \label{fig:quantum_noise}
    }
    \vspace{-10pt}
    \caption{(a) Classical simulation has unscalable computational and memory costs. (b) Noises create significant accuracy gaps between \pqc (\qnn) classical simulation and on-chip training.
    (c) Small gradients suffer from larger relative errors, thus being less reliable.}
    \vspace{-15pt}
    \label{fig:motivation}
\end{figure}

In order to achieve \pqc quantum advantage, the number of qubit needs to be large enough, which casts great difficulty in the parameter training process. In existing \pqc work~\citep{harrow2009quantum, lloyd2013quantum}, the primary focus has been building quantum models that can outperform classical model accuracy. Thus they typically perform training on classical computers through software simulations and then perform inference with simulators as well (Figure~\ref{fig:teaser} top). Although classical simulation is useful in understanding the capabilities of small-size \pqc, it is not scalable due to the exponentially increased time and memory costs ($\mathcal{O}(2^{n}), n$ is the qubit number). As shown in Figure~\ref{fig:theoretical_cost}, the space (\#Regs) and time (\#Ops) complexity of classical simulation grow exponentially as the number of qubits increases. To the authors' knowledge, this is the \emph{first experimental demonstration} of efficient and scalable \pqc on-chip training protocol.
The optimization of parametrized quantum gates is offloaded to the quantum chips with \textit{in-situ} gradient computation using \emph{parameter shift}~\citep{mitarai2018quantum}. We also perform \pqc evaluation on real quantum machines, making the results more \emph{practical} as in Figure~\ref{fig:teaser} bottom.

One of the major challenges to enable scalable and efficient \pqc on-chip learning is the robustness against quantum noise. In the current Noisy Intermediate Scale Quantum (NISQ)~\citep{preskill2018quantum} era, the gate error rates on real quantum devices are non-negligible ($10^{-3}$ to $10^{-2}$). In the context of \pqc, such errors will lead to \emph{noisy gradients} which can slow down convergence or even make training unstable.
As shown in Figure~\ref{fig:classical_quantum_gap}, large gaps exist between the quantum on-chip training results and the classical noise-free simulation results.

By carefully investigating the on-chip training process, we observe that small gradients tend to have large relative variations or even wrong directions under quantum noises, as shown in Figure~\ref{fig:quantum_noise}.
Also, not all gradient computations are necessary for the training process, especially for small-magnitude gradients. 
Those observations provide great opportunities for us to boost the robustness and efficiency of \pqc on-chip learning. Inspired by that, we propose a \emph{probabilistic gradient pruning} method to predict and only compute gradients of high reliability. 
Hence we can reduce noise impact and also save the required number of circuit runs on real quantum machines. In this paper, we are mainly using \textit{QNNs as benchmarks} but the techniques can also be applied to \textit{other PQCs} such as Variational Quantum Eigensolver (VQE). \name has following contributions:
\begin{itemize}[leftmargin=*]
\setlength{\itemindent}{0.5em}
    \item We are the \emph{first work} to demonstrate the practicality of parameter shift on \emph{NISQ machines}, achieving high \pqc learning accuracy.
    \item A probabilistic gradient pruning method is proposed to improve the noise robustness by 5-7\% and reduce the number of inference on real QC by 2$\times$ while maintaining the accuracy.
    \item Experimental deployment of QNN on 5 real quantum machines demonstrates that the proposed method can achieve over \textbf{90\%} and \textbf{60\%} accuracy for 2-class and 4-class image recognition tasks.
    Our framework enables scalable, robust, and efficient training of PQCs with large number of qubits and parameters.
    \item We open-source the parameter shift on-chip \pqc training and gradient pruning code in the \href{https://github.com/mit-han-lab/torchquantum}{\torchquantum} library.
\end{itemize}

\section{Background}

\noindent\textbf{Quantum basics.}
Quantum circuits use quantum bit (called \emph{qubit}) to store information, which is a linear combination of two basis states: $\ket{\psi}=\alpha\ket{0}+\beta\ket{1}$, for $\alpha,\beta\in\mathbb{C}$, satisfying $|\alpha|^2+|\beta|^2=1$.
An $n$-qubit system can represent a linear combination of 2$^n$ basis states.
A 2$^n$-length complex statevector of all combination coefficients is used to describe the quantum state. To perform computation on a quantum system, a sequence of parametrized quantum gates are applied to perform unitary transformation on the statevector, i.e., $\ket{\psi(\boldsymbol{x}, \boldsymbol{\theta})}=\cdots U_2(\boldsymbol{x},\theta_2)U_1(\boldsymbol{x},\theta_1)\ket{0}$, where $\boldsymbol{x}$ is the input data, and $\boldsymbol{\theta}=(\theta_1, \theta_2, \dotsc)$ are trainable parameters in quantum gates. In this way, input data and trainable parameters are embedded in the quantum state $\ket{\psi(\boldsymbol{x}, \boldsymbol{\theta})}$.
The computation results are obtained by qubit readout which measures the probability of a qubit state $\ket{\psi}$ collapsing to either $\ket{0}$ (i.e., output $y=+1$) or $\ket{1}$ (i.e., output $y=-1$) according to $|\alpha|^2$ and $|\beta|^2$. With sufficient samples, we can compute the expectation value: $\E[y] = (+1)|\alpha|^2 + (-1)|\beta|^2$.
A non-linear network can be constructed to perform ML tasks by cascading multiple blocks of quantum gates and measurements.

\noindent\textbf{Quantum noise.}
In real quantum computer systems, errors (noises) would occur due to unwanted interactions between qubits, imperfect control signals, or interference from the environment~\citep{hsieh2020realistic}. For example, quantum gates introduce \emph{operation errors} (e.g., coherent errors and stochastic errors) into the system, and qubits also suffer from \emph{decoherence error} (spontaneous loss of its stored information) over time.
These noisy systems need to be characterized~\citep{magesan2012characterizing} and calibrated~\citep{ibm_2021} frequently to mitigate the noise impact. 

\noindent\textbf{Quantum neural networks.}
Quantum Machine Learning (QML)~\citep{biamonte2017quantum, wang2021quantumnas, wang2021roqnn, liang2021can, wang2021exploration} aims to leverage QC techniques to solve machine learning tasks and achieve much higher efficiency. 
The path to \emph{quantum advantage} on QML is typically provided by the quantum circuit's ability to generate and estimate highly complex kernels ~\citep{havlivcek2019supervised}, which would otherwise be intractable to compute with conventional computers.
They have been shown to have potential speed-up over classical counterparts in various tasks, including metric learning ~\citep{lloyd2020quantum}, data 
\begin{figure}
\vspace{-15pt}
    \centering
    \includegraphics[width=\columnwidth]{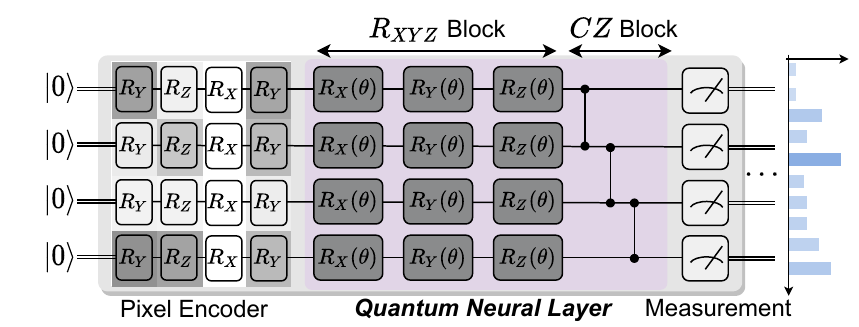}
    \caption{Quantum Neural Network (QNN) architecture.}
    \label{fig:qnn}
    \vspace{-15pt}
\end{figure} analysis~\citep{lloyd2016quantum}. 
As shown in Figure~\ref{fig:qnn}, the quantum neural network is one type of QML model using variational quantum circuits with trainable parameters to accomplish feature encoding of input data and perform complex-valued linear transformations thereafter. 
Most of QNN trainings are exploratory and rely on classical simulation of small quantum systems. In our work, on the contrary, we explore the practical setting: the \qnn training and inference are both performed on real quantum devices.

\noindent\textbf{Pruning.} Pruning techniques are widely used in the field of DNN~\citep{han2015deep, wang2020apq, wang2021spatten, zhang2020sparch}, performing an important role of the trade-off between accuracy and memory or time cost~\citep{nguyenmeidine2020progressive}. Recently, pruning techniques have been used in quantum tasks. Pruning the ansatz can bring time-efficient circuit and even higher performance on real QC~\citep{wang2021quantumnas}. In our work, we apply pruning techniques to prune unreliable gradients in order to mitigate the noise during training.

\section{Methodology}
To enable \pqc on-chip learning, we first introduce an \textit{in-situ} quantum gradient computation via parameter shift and its real QC implementation.
A probabilistic gradient pruning method is proposed to save the gradient computation cost with enhanced noise-robustness and training efficiency. We study \qnn as the benchmark \pqc.

\subsection{Parameter Shift Rule for Quantum Gradients}

Parameter shift rule states that we can calculate the gradient of each parameter in some quantum circuits by simply shifting the parameter twice and calculating the difference between two outputs, without changing the structure of circuits or using any ancilla qubits. 
Prior works elaborate it based on quantum circuit function~\citep{crooks2019gradients}, however, in the next subsection we will show how parameter shift rules combined with backpropagation can be used in a real \pqc task.
Suppose an $m$-qubit quantum circuit is parametrized by $n$ parameters $\theta=[\theta_1,\cdots,\theta_i,\cdots,\theta_n]$, the expectation value of measurements of this circuit can be represented by a \textbf{circuit function},
\begin{equation}
    \small
    \label{eq:QuantumCircuit}
    f(\theta)=\langle\psi|U(\theta_i)^{\dag}\widehat{Q}U(\theta_i)|\psi\rangle, \quad f(\theta)\in\mathbb{R}^{m}, \theta\in\mathbb{R}^n.
\end{equation}
where $\theta_i$ is the scalar parameter whose gradient is to be calculated, and $U(\theta_i)$ is the gate where $\theta_i$ lies in. 
Here, for notation simplicity, we have already absorbed the unitaries before $U(\theta_i)$ into 
$\langle\psi|$, $|\psi\rangle$.
Unitaries after $U(\theta_i)$ and observables are fused into $\widehat{Q}$. Usually, the gates used in \pqc can be written in the form $U(\theta_i)=e^{-\frac{i}{2}\theta_i H}$. Here $H$ is the Hermitian generator of $U$ with only 2 unique eigenvalues +1 and -1 ($H$'s eigenvalues can be $\pm r$, but for simplicity we assume it's $\pm 1$). In this way, the gradients of the circuit function $f$ with respect to $\theta_i$ are,
\begin{equation}
    \small
    \label{eq:ParamShift}
    \begin{aligned}
    &\frac{\partial f(\theta)}{\partial \theta_i}=\frac{1}{2}\Big(f\big(\theta_+\big)-f\big(\theta_{-}\big)\Big), \\ &\theta_+=[\theta_1,\cdots,\theta_i+\frac{\pi}{2},\cdots,\theta_n], \theta_{-}=[\theta_1,\cdots,\theta_i-\frac{\pi}{2},\cdots,\theta_n],
    \end{aligned}
\end{equation}
where $\theta_+$ and $\theta_{-}$ are the \emph{positive shift} and \emph{negative shift} of $\theta$.
Note that this parameter shift rule is\emph{ fundamentally different} from any numerical difference methods that only approximate the directional derivatives.
Instead, Eq.~\ref{eq:ParamShift} calculates the \emph{exact} gradient w.r.t $\theta_i$ without any approximation errors or numerical issues. 

We apply \texttt{softmax} on the expectation values of measurements $f(\theta)$ as the predicted probability for each class. 
Then we calculate the cross entropy between the predicted probability distribution $p$ and the target distribution $t$ as the classification loss $\mathcal{L}$, 
\begin{equation}
    \small
    \label{eq:CrossEntropy}
    \mathcal{L}(\theta)=-t^T\cdot\texttt{softmax}(f(\theta))=-\sum_{j=1}^m t_j \log{p_j},\quad p_j=\frac{e^{f_j(\theta)}}{\sum_{j=1}^m e^{f_j(\theta)}}.
\end{equation}
Then the gradient of the loss function with respect to $\theta_i$ is $\frac{\partial\mathcal{L}(\theta)}{\partial \theta_i}=\big(\frac{\partial\mathcal{L}(\theta)}{\partial f(\theta)}\big)^T\frac{\partial f(\theta)}{\partial \theta_i}$.

Here $\frac{\partial f(\theta)}{\partial \theta_i}$ can be calculated on real quantum circuit by the parameter shift rule, and $\frac{\partial\mathcal{L}(\theta)}{\partial f(\theta)}$ can be efficiently calculated on classical devices using backpropagation supported by automatic differentiation frameworks, e.g., PyTorch and TensorFlow.

Now we derive the parameter shift rule used in our \pqc models.

Assume $U(\theta_i)=R_X(\theta_i),R_X(\alpha)=e^{-\frac{i}{2}\alpha X}$, where $X$ is the Pauli-X matrix.

Firstly, the RX gate is,
\begin{equation}
    \small
    \label{eq:RXGate}
    \begin{aligned}
R_X(\alpha)&=e^{-\frac{i}{2}\alpha X}=\sum_{k=0}^{\infty}(-i\alpha/2)^kX^k/k!\\
&=\sum_{k=0}^{\infty}(-i\alpha/2)^{2k}X^{2k}/(2k)!+\sum_{k=0}^{\infty}(-i\alpha/2)^{2k+1}X^{2k+1}/(2k+1)!\\
&=\sum_{k=0}^{\infty}(-1)^k(\alpha/2)^{2k}I/(2k)!-i\sum_{k=0}^{\infty}(-1)^k(\alpha/2)^{2k+1}X/(2k+1)!\\
&=\cos(\alpha/2)I-i\sin(\alpha/2)X.
\end{aligned}
\end{equation}
Let $\alpha=\frac{\pi}{2}$, $R_X(\pm\frac{\pi}{2})=\frac{1}{\sqrt{2}}(I\mp iX)$.

As $f(\theta)=\langle\psi|R_X(\theta_i)^{\dag}\widehat{Q}R_X(\theta_i)|\psi\rangle$, $R_X(\alpha)R_X(\beta)=R_X(\alpha+\beta)$, and $\frac{\partial}{\partial \alpha}R_X(\alpha)=-\frac{i}{2}XR_X(\alpha)$,
we have
\begin{equation}
\small
\label{eq:ParamShiftDerivation}
\begin{aligned}
\frac{\partial f(\theta)}{\partial \theta_i}
=&\langle\psi|R_X(\theta_i)^{\dag}(-\frac{i}{2}X)^{\dag}\widehat{Q}R_X(\theta_i)|\psi\rangle+\langle\psi|R_X(\theta_i)^{\dag}\widehat{Q}(-\frac{i}{2}X)R_X(\theta_i)|\psi\rangle\\
=&\frac{1}{4}(\langle\psi|R_X(\theta_i)^{\dag}(I-iX)^{\dag}\widehat{Q}(I-iX)R_X(\theta_i)|\psi\rangle\\&-\langle\psi|R_X(\theta_i)^{\dag}(I+iX)^{\dag}\widehat{Q}(I+iX)R_X(\theta_i)|\psi\rangle)\\
=&\frac{1}{2}(\langle\psi|R_X(\theta_i)^{\dag}R_X(\frac{\pi}{2})^{\dag}\widehat{Q}R_X(\frac{\pi}{2})R_X(\theta_i)|\psi\rangle\\&-\langle\psi|R_X(\theta_i)^{\dag}R_X(-\frac{\pi}{2})^{\dag}\widehat{Q}R_X(-\frac{\pi}{2})R_X(\theta_i)|\psi\rangle)\\
=&\frac{1}{2}(f(\theta_+)-f(\theta_-)).
\end{aligned}
\end{equation}

Without loss of generality, the derivation holds for all unitaries of the form $e^{-\frac{i}{2}\alpha H}$, e.g., RX, RY, RZ, XX, YY, ZZ, where $H$ is a Hermitian matrix with only 2 unique eigenvalues +1 and -1.

In our circuit functions, we assume each parameter lies in exactly one gate. 
However, there are cases that one parameter lies in multiple gates. 
In that case, we only need to calculate the gradient of the parameter in those gates separately and sum the gradients up to get the gradient of that parameter.
\begin{figure}
    \centering
    \includegraphics[width=\columnwidth]{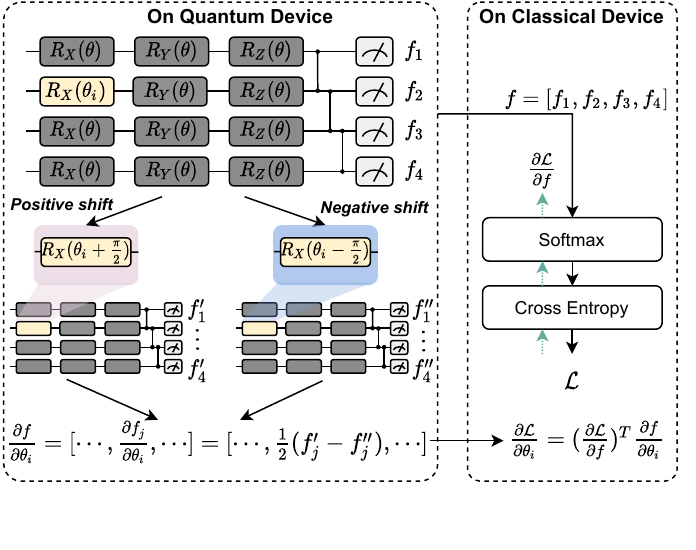}
    \vspace{-40pt}
    \caption{Quantum gradient calculation using the parameter shift rule on real quantum devices.}
    \vspace{-20pt}
    \label{fig:paramshift}
\end{figure}

\subsection{\textit{In-situ} Gradient Computation on Real QC}
To realize \pqc on-chip learning, we implement a TrainingEngine, described in Alg.~\ref{alg:qnn_train_alg}. 
This TrainingEngine contains three parts.

\noindent\textbf{Jacobian calculation via parameter shift.}~
In the first part, we sample a mini-batch of training data $\mathcal{I}$ in Line 6. 
For each example of the mini-batch, we set up the quantum encoder gates and then iteratively evaluate gradients for all parameters. 
In each iteration, we shift the parameter $\theta_i$ twice by $+\pi/2$ and $-\pi/2$ respectively. 
After each shift, we execute the shifted circuit on quantum hardware. 
The circuit will be \emph{created, validated, queued, and finally run on real quantum machines}. 
As soon as we get the returned results of the two shifted circuits, i.e., $f(\theta_+)$ and $f(\theta_-)$, we apply Eq.~\ref{eq:ParamShift} to obtain the upstream gradient $\frac{\partial f(\theta)}{\partial\theta_i}$, illustrated in the left part of Figure~\ref{fig:paramshift}.
Finally, we obtain the Jacobian matrix $\frac{\partial f(\theta)}{\partial\theta}$.

\begin{figure*}
    \centering
    \includegraphics[width=0.98\textwidth]{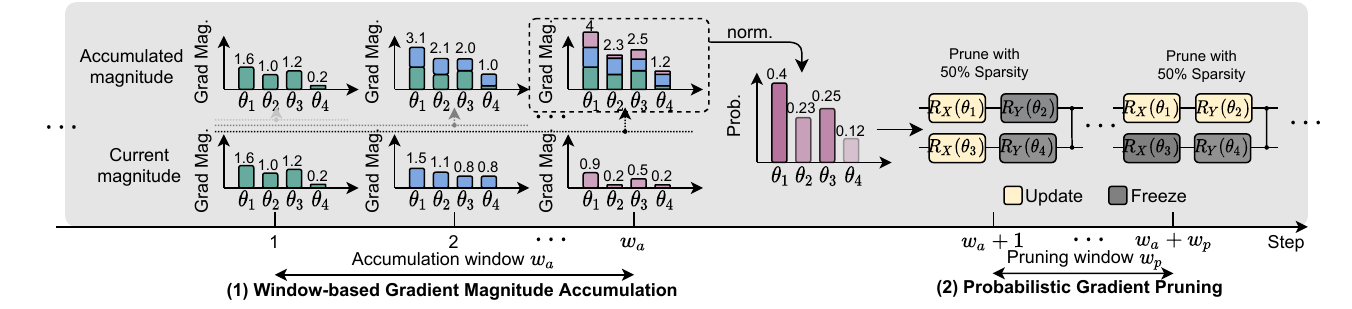}
    \caption{Efficient on-chip quantum gradient calculation with probabilistic gradient pruning.
    Gradient magnitudes are accumulated within the accumulation window and used as the sampling distribution.
    Based on the distribution, gradients are probabilistically pruned with a ratio $r$ in the pruning window to mitigate noises and stabilize training.
    }
    \label{fig:gradient_pruning}
\end{figure*}

\noindent\textbf{Down-stream gradient backpropagation.}~
In the second part, we run the circuit without shift and get the measurement result $f(\theta)$. 
Then we apply \texttt{softmax} and cross-entropy function to the measured logits. 
In the end, we get the training loss $\mathcal{L}(\theta)$. 
Then we run back-propagation \emph{only from the loss to the logits} to get the down-stream gradients $\frac{\partial\mathcal{L}(\theta)}{\partial f(\theta)}$, shown in the right part of Figure~\ref{fig:paramshift}.

\noindent\textbf{Gradient calculation.}~
In the third part, we calculate the dot-product between down-stream gradients and the Jacobian and get the final gradients $\frac{\partial\mathcal{L}(\theta)}{\partial\theta}=(\frac{\partial f(\theta)}{\partial\theta})^T\frac{\partial\mathcal{L}(\theta)}{\partial f(\theta)}$.

\subsection{Probabilistic Quantum Gradient Pruning}
\label{sec:GradientPruning}

On quantum chips, there exist various noises and errors that could potentially diminish the fidelity of the computation results.
When the gradient magnitude is small, noises could easily overwhelm the signals, such that the \emph{gradients calculated on real quantum circuit become unreliable when they have small magnitude}. 
Those unreliable gradients have harmful effects on training convergence. 
Skipping the evaluation on those unreliable gradients can benefit both training convergence and efficiency. 
Besides, we observe that for most parameters, if the gradient magnitudes are far from zero for several steps, it will likely keep far from zero in the next several steps. 
Similarly, if the gradient magnitude remains small for some steps, it will likely keep small in the next several steps. 
This means the \emph{gradient reliability is predictable} to some extent. 
Therefore, we propose the gradient pruning method to sample the parameters whose gradients are more reliable. 
This method helps training converge faster while also saving time by skipping the evaluation of unreliable gradients.

Alg.~\ref{alg:qnn_train_alg} describes the \pqc on-chip training flow with probabilistic gradient pruning.
We divide all the training steps into $S$ stages and perform the pruning method periodically on each stage. 
For every stage, we split it into two phases, shown in Figure~\ref{fig:gradient_pruning}. 
The first phase is called \emph{magnitude accumulation} with an accumulation window width $w_a$, and the second is called \emph{probabilistic gradient pruning} (PGP) with a pruning window width $w_p$. 
We only apply pruning in the second phase, while the parameter subset is sampled from a probability distribution $\tilde{\theta}= \{\theta_i\sim P_M(\theta)|1\leq i\leq (1-r)n\}$ based on the gradient information collected within the accumulation window.

In Lines 4-9, within the accumulation window, we record the magnitude of gradients of each parameter in each step and accumulate them until the window is over. 
At the end of the first phase, we can get an accumulator $M$ that records the accumulated gradient magnitude for each parameter. 
Thus, when the pruning phase starts, we normalize the accumulated gradient magnitude and pass it to our sampler as the sampling distribution. 
In each pruning step, the sampler samples a subset of parameters $\tilde{\theta}$ with a pruning ratio of $r$, and we only evaluate gradients for them while the rest $\theta\backslash\tilde{\theta}$ is temporarily frozen.

There are three important hyper-parameters in our gradient pruning method: 1) accumulation window width $w_a$, 2) pruning ratio $r$, and 3) pruning window width $w_p$. 
The accumulation window width and pruning window width decide the reliability of the gradient trend evaluation and our confidence in it, respectively. 
The pruning ratio can be tuned to balance the gradient variances caused by noise perturbation and pruning. 
Thus, the percentage of the time saved by our probabilistic gradient pruning method is $r\frac{w_p}{w_a+w_p}\times100\%$.
In our experiments, we find that the setting 
($w_a$=1, $w_p$=2$\sim$3, $r$=0.3$\sim$0.5)
usually works well in all cases.

\begin{algorithm2e}[t]
    \SetAlgoLined
    \SetKwInOut{Input}{Input}
    \SetKwInOut{Output}{Output}
    
    \Input{ Accumulation window width $w_a$, gradient pruning ratio $r$, pruning window width $w_p$, training objective $\mathcal{L}$, initial parameters $\theta^0\in\mathbb{R}^n$, training data $\mathcal{D}_{trn}$, initial step size $\eta^0$, and total stages $S$.
    }
    $\theta\gets\theta^0,~~\eta\gets\eta^0~~t\gets0$\;
    \For{$s=1,2,.\cdots,S$}{
    \text{Initialize gradient magnitude accumulator }$M\gets \mathbf{0}^n$\;
    \For{$\tau_a=1,2,\cdots,w_a$}{
     $t\gets t+1$\;
     \text{Sample a mini-batch }$\mathcal{I}\sim\mathcal{D}_{trn}$\;
     \text{\textit{In-situ} gradient evaluation via parameter shift} $\nabla_{\theta}\mathcal{L}_{\mathcal{I}}(\theta)=\frac{1}{2}(\frac{\partial f(\theta)}{\partial \theta})^T\frac{\partial\mathcal{L}(\theta)}{f(\theta)}$\;
     \text{Parameter update: }$\theta\gets\theta-\eta\nabla_{\theta}\mathcal{L}_{\mathcal{I}}(\theta)$\;
     \text{Update magnitude accumulator }$M\gets M+|\nabla_{\theta}\mathcal{L}_{\mathcal{I}}(\theta)|$\;
     }
     \For{$\tau_p \gets 1,2,\cdots,w_p$}{
     $t\gets t+1$\;
     \text{Sample a mini-batch }$\mathcal{I}\sim\mathcal{D}_{trn}$\;
     Sample a subset with a ratio $r$ based on accumulated gradient magnitude:
     $\tilde{\theta}= \{\theta_i\sim P_M(\theta)|1\leq i\leq (1-r)n\}$\;
     $\tilde{\theta}\gets\tilde{\theta}-\eta\nabla_{\tilde{\theta}}\mathcal{L}_{\mathcal{I}}(\theta)$\;
     }
     }
    \Output{Converged parameters $\theta$}
    \caption{\pqc On-Chip Training with Probabilistic Gradient Pruning}
    \label{alg:qnn_train_alg}
\end{algorithm2e}

\section{Experiments}
\label{sec:experiments}
In this section, we deploy our \pqc on-chip learning framework on real QC and evaluate it on 5 \qnn tasks for image and vowel recognition.
Compared with classical QNN training protocols, we can achieve 2-4\% real QC test accuracy improvement with 2$\times$ convergence speedup.
We also conduct extensive ablation studies to validate our scalability and the effectiveness of the proposed probabilistic gradient pruning method.
\subsection{Experiment Setups}

\textbf{Benchmarks}. We conduct our experiments on 5 QML tasks. QML are all classification tasks including MNIST~\citep{726791} 4-class (0, 1, 2, 3), 2-class (3 and 6); Fashion~\citep{xiao2017fashion} 4-class (t-shirt/top, trouser, pullover,
dress), 2-class (dress and shirt); Vowel 4-class(hid, hId, had, hOd).
MNIST and Fashion 2-class use the front 500 images as the training set and randomly sampled 300 images as the validation set. MNIST, Fashion 4-class uses the front 100 images as the training set and also randomly sampled 300 images as the validation set. The input images are all $28\times 28$. We firstly center-crop them to $24\times 24$ and then down-sample them to $4\times 4$ for MNIST and Fashion 2 and 4-class tasks. Vowel 4-class uses the front 100 samples as the training set and randomly sampled 300 samples as the validation set. For each sample, we perform principal component analysis (PCA) for the vowel features and take the 10 most significant dimensions.

\begin{table}[t]
\centering

\renewcommand*{\arraystretch}{0.8}
\setlength{\tabcolsep}{1pt}

\caption{Accuracy comparison among different settings.
"Simu." represents "simulation".}
\vspace{-10pt}
\label{tab:main_results}
\resizebox{0.48\textwidth}{!}{%
\begin{tabular}{lcccccc}
\toprule
Method & Acc. & MNIST-4 & MNIST-2 & Fashion-4 & Fashion-2 & Vowel-4 \\
& & Jarkata & Jarkata & Manila & Santiago & Lima \\
\midrule
\midrule
Classical-Train & Simu. & 0.61 & 0.88 & 0.73 & 0.89 & 0.37 \\
\midrule
Classical-Train & & 0.59 & 0.79 & 0.54 & 0.89 & 0.31 \\
QC-Train & QC & 0.59 & 0.83 & 0.49 & 0.84 & 0.34 \\ 
\textbf{QC-Train-PGP} && \textbf{0.64} & \textbf{0.86} & \textbf{0.57} & \textbf{0.91} & \textbf{0.36} \\ 
\bottomrule
\end{tabular}%
}
\vspace{-20pt}
\end{table}
\begin{figure}
    \centering
    \subfloat[]{
    \includegraphics[width=0.23\textwidth]{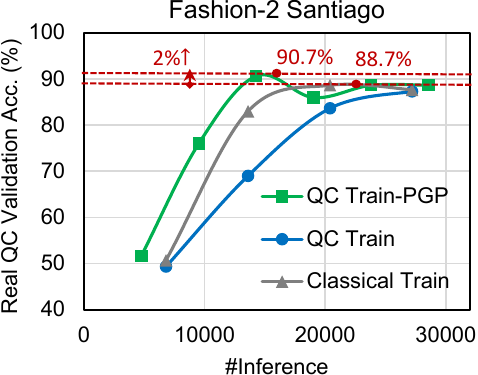}
    \label{fig:fashion_train_curve}
    }
    \subfloat[]{
    \includegraphics[width=0.235\textwidth]{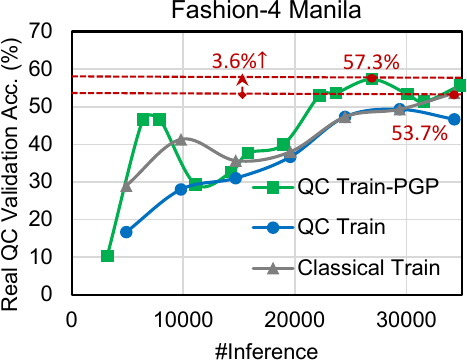}
    \label{fig:fashion4_train_curve}
    }
    \vspace{-10pt}
    \caption{Real QC validation accuracy curves on different datasets and different quantum devices.}
    \vspace{-10pt}
    \label{fig:train_curve}
\end{figure}
All the tasks use four logical qubits. To embed classical image and vowel features to the quantum states, we first flatten them and then encode them with rotation gates. For down-sampled $4\times 4$ images, we use 4RY, 4RZ, 4RX, and 4RY gates as the encoder. We put the 16 classical input values to the phases of 16 rotation gates, respectively. Therefore we can encode the classical values to quantum states. For 10 vowel features, we use 4RY, 4RZ, and 2RX gates for encoding.

The encoding gates are our hand-designed circuits. Our circuits are composed of several layers. 
There are 7 kinds of layers used to construct our circuits. ($i$) RX layer: Add RX gates to all wires; ($ii$) RY layer: same structure as in RX layer; ($iii$) RZ layer: same structure as in RX layer; ($iv$) RZZ layer: add RZZ gates to all logical adjacent wires and the logical farthest wires to form a ring connection, for example, an RZZ layer in a 4-qubit circuit contains 4 RZZ gates which lie on wires 1 and 2, 2 and 3, 3 and 4, 4 and 1; ($v$) RXX layer: same structure as in RZZ layer; ($vi$) RZX layer: same structure as in RZZ layer; ($vii$) CZ layer: add CZ gates to all logical adjacent wires. 

For MNIST and Fashion 2-class tasks, the circuit contains 1 RZZ layer followed by 1 RY layer. For MNIST 4-class task, the circuit contains 3 RX+RY+RZ+CZ layers (1 RX layer, 1 RY layer, 1 RZ layer, and 1 CZ layer in series). For Fashion 4-class task, the circuit contains 3 RZZ+RY layers (1 RZZ layer followed by 1 RY layer). For Vowel 4-class task, the circuit contains 2 RZZ+RXX layers (1 RZZ layer followed by 1 RXX layer).

For the output of our quantum circuits, we measure the expectation values on Pauli-Z basis and obtain a value [-1, 1] from each qubit. For 2-class, we sum the qubit 0 and 1, 2, and 3 respectively to get 2 output values. For 4-class, we just use the four expectation values as 4 output values. Then we process the output values by Softmax to get probabilities.

\noindent\textbf{Quantum devices and compiler configurations.}~
We use IBM quantum computers via qiskit API~\citep{ibm_2021} to submit our circuits to real superconducting quantum devices and achieve quantum on-chip training. 
We set all the circuits to run 1024 shots.

\noindent\textbf{Baseline.}~
We have two baselines. (1) QC-Train: We train our model without gradient pruning, i.e., calculating gradients of every parameter in each step. 
The gradient calculation is deployed on real quantum circuits. 
(2) Classical-Train: We train our QNN model completely on classical computers. 
We use a vector to record the amplitudes of the quantum state, utilize complex matrix multiplication to simulate quantum gates, and sample based on the amplitude vector to simulate quantum measurement.

The QC-Train-PGP line shows training on real quantum circuits while applying our probabilistic gradient pruning. In all the cases, we adopt accumulation window size 1, pruning ratio 0.5, and pruning window size 2, except for Fashion-4, we adopt pruning ratio 0.7, and other settings remain the same.

\subsection{Main Results}

\noindent\textbf{QNN results.}~
Table~\ref{tab:main_results} shows the accuracy of comparison on 5 tasks. 
In each task, we show 4 accuracy values, which are (1) accuracy of Classical-Train tested on classical devices,  (2) accuracy of Classical-Train tested on real quantum circuits; (3) accuracy of QC-Train tested on real quantum circuits; (4) accuracy of QC-Train-PGP tested on real quantum circuits. In each task, the accuracy is collected after finishing a certain number of circuit runs. We train and evaluate MNIST-2 and MNIST-2 on ibmq\_jakarta, Fashion-4 on ibmq\_manila, Fashion-2 on ibmq\_santiago, and Vowel-4 on ibmq\_lima.

The noise-free accuracy is usually the highest among the other three, because it represents the accuracy without any noise perturbation. 
The QC-Train-PGP usually takes second place because compared to Classical-Train, it has the advantage of noise awareness, and compared to QC-Train, it suffers \textit{less} from noise thanks to gradient pruning.

\begin{figure}
    \centering
    \includegraphics[width=\columnwidth]{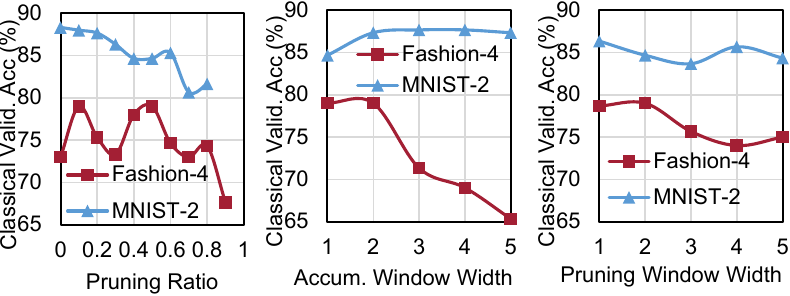}
    \vspace{-20pt}
    \caption{Ablation on pruning ratio, accumulation window width, and pruning window width.}
    \vspace{-15pt}
    \label{fig:ablation}
\end{figure}

\noindent\textbf{Training curves.}~ Figure~\ref{fig:train_curve} shows the real QC validation accuracy curve during training.
The X-axis is the number of inferences (how many circuits have been run). 
The Y-axis is the accuracy of the validation dataset tested on real quantum circuits. 
MNIST 4-class runs on the ibmq\_jakarta machine. 
We observe that given a fixed inference budget, our QC-Train-PGP achieves the best accuracy of 63.7\% while the Classical-Train only achieves 59.3\%.

We further train Fashion 2-class on ibmq\_santiago. 
QC-Train-PGP only takes 13.9k inferences to reach the peak accuracy 90.7\%, while the best accuracy Classical-Train can achieve is merely 88.7\% at the cost of over 30k inferences. 

\subsection{Ablation Studies}

\noindent\textbf{Ablation on gradient pruning.}~
In Figure~\ref{fig:ablation}, we evaluate the training performance with different pruning ratios $r$, accumulation window size $w_a$, and pruning window size $w_p$ on Fashion-4 and MNIST-2 tasks. 
We find that the $r=0.5$ is generally a good setting for our tasks.
Overly large pruning ratios will induce too many gradient variances that harm the training convergence.
For the accumulation window size, $w_a=1$ or $2$ are suitable choices.
When $w_a$ is too large, the accumulated gradient magnitudes are similar among all parameters, leading to a nearly uniform sampling distribution. 
This will bring undifferentiated pruning, and the accuracy will drop as the Fashion-4 curve shows. 
The pruning window $w_p$ should also not be too large. 
As $w_p$ grows, the accumulated gradient magnitudes used to instruct our pruning become less reliable.

\begin{table}[t]

\centering
\captionof{table}{The proposed probabilistic pruning is better than deterministic pruning.}
\vspace{-10pt}
\resizebox{\columnwidth}{!}{%
\begin{tabular}{lcccc}
\toprule
Method & MNIST-4 & MNIST-2 & Fashion-4 & Fashion-2 \\
\midrule
Deterministic & 0.61 & 0.82 & 0.72 & 0.89 \\
Probabilistic & \textbf{0.62} & \textbf{0.85} & \textbf{0.79} & \textbf{0.90} \\
\bottomrule
\end{tabular}%
\label{tab:deterministic}
}
\vspace{-10pt}
\end{table}

\begin{figure}
    \centering
    \includegraphics[width=\columnwidth]{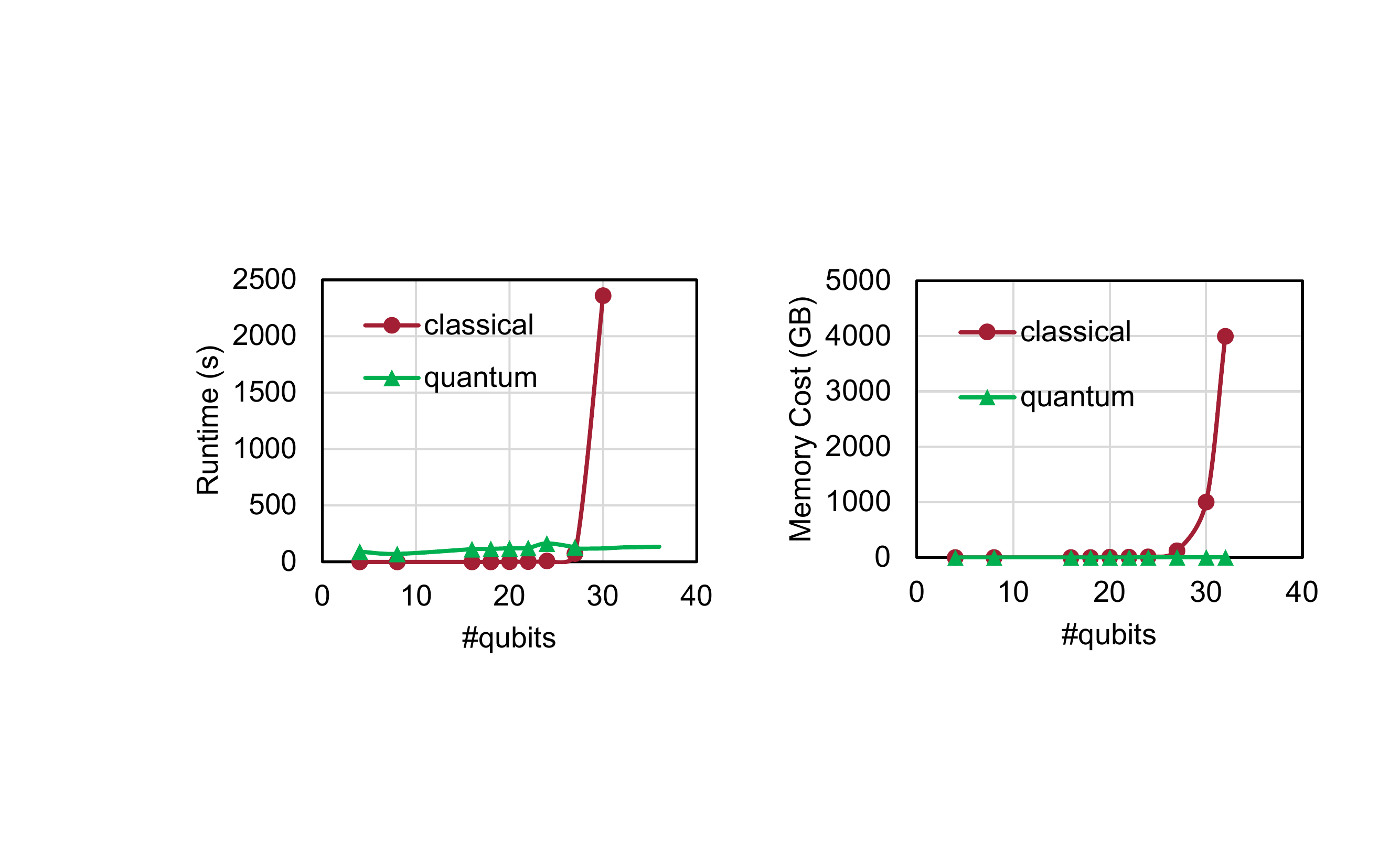}
    \vspace{-20pt}
    \caption{Runtime and memory cost comparison between classical simulation and quantum on-chip run.}
    \vspace{-15pt}
    \label{fig:scalability}
\end{figure}

\begin{table}
\centering
\caption{Adam optimizer can outperform SGD and Momentum optimizers.}
\vspace{-10pt}
\label{tab:optimizer}
\resizebox{\columnwidth}{!}{%
\begin{tabular}{lcccc}
\toprule
Optimizer & MNIST-4 & MNIST-2 & Fashion-4 & Fashion-2 \\
\midrule
SGD & 0.5 & 0.8 & 0.45 & 76 \\
Momentum & 0.55 & 0.83 & 0.66 & 0.90 \\
Adam & \textbf{0.61} & \textbf{0.88} & \textbf{0.75} & \textbf{0.91} \\
\bottomrule
\end{tabular}%

}

\vspace{-15pt}
\end{table}
\noindent\textbf{Discussion on scalability.}~
Figure~\ref{fig:scalability} shows the superior scalability of quantum on-chip training.
Classical simulation runtime exponentially increases as \#qubits scales up, 
while the runtime on real quantum machines scales nearly linearly to \#qubits. The classical curve in Figure~\ref{fig:scalability} represents runtime and memory cost of running 50 circuits of different \#qubits with 16 rotation gates and 32 RZZ gates. The curve before 22 qubits is measured on a single NVIDIA RTX 2080 Ti GPU; points after 24 qubits are extrapolated. 
The quantum curve before 27 qubits is tested on ibmq\_toronto; the points after 30 qubits are extrapolated.

We can observe clear quantum advantages on circuits with more than 27 qubits.
In terms of memory cost, classical simulation consumes thousands of Gigabits for storage which is intractable.
In contrast, on quantum machines, the information is stored in the quantum state of the circuit itself with negligible memory cost.

\noindent\textbf{Probabilistic vs. deterministic gradient pruning.} 
Our pruning is decided by a random sampler based on the accumulated gradient magnitude. We call this probabilistic pruning. If the sampler only samples the parameters with the biggest accumulated gradient magnitude, this is called deterministic pruning. We adopt probabilistic pruning instead of deterministic pruning because deterministic pruning limits the degree of freedom and increases the gradient sampling bias.
Table~\ref{tab:deterministic} shows that deterministic pruning has 1\%-7\% accuracy loss compared with probabilistic pruning.

\noindent\textbf{Different optimizers.}~ 
Table~\ref{tab:optimizer} shows the accuracy tested on classical devices trained with different optimizers. 
The learning rate is controlled by a cosine scheduler from 0.3 in the beginning to 0.03 in the end.
We test SGD, SGD with a momentum factor of 0.8, and Adam on MNIST-4, MNIST-2, Fashion-4, and Fashion-2, and found that Adam always performs the best. 
Hence, all the experiments are done using Adam optimizers by default. 

\section{Conclusion}
In this work, for the first time, we present an efficient and robust on-chip training framework for \pqc
and demonstrate its effectiveness on real quantum devices.
By leveraging parameter shift, we can calculate the exact quantum gradients directly on quantum machines, thus achieving high \emph{scalability}. To alleviate the negative impact of quantum noises on gradients, we further propose the probabilistic gradient pruning technique to avoid updating parameters with unreliable gradients. 
Experimental results on 5 classification tasks and 5 machines demonstrate that \name achieves comparable accuracy with noise-free simulations. We hope \name can open an avenue towards practical training of large \pqc models for quantum advantage.

\section*{Acknowledgment}

\small
We acknowledge NSF CAREER Award \#1943349, MIT-IBM Watson AI Lab, Baidu Fellowship, Qualcomm Innovation Fellowship, and IBM Quantum.

{\small
\balance
\bibliographystyle{ACM-Reference-Format}
\bibliography{main.bib}


\begin{thebibliography}{23}


\ifx \showCODEN    \undefined \def \showCODEN     #1{\unskip}     \fi
\ifx \showDOI      \undefined \def \showDOI       #1{#1}\fi
\ifx \showISBNx    \undefined \def \showISBNx     #1{\unskip}     \fi
\ifx \showISBNxiii \undefined \def \showISBNxiii  #1{\unskip}     \fi
\ifx \showISSN     \undefined \def \showISSN      #1{\unskip}     \fi
\ifx \showLCCN     \undefined \def \showLCCN      #1{\unskip}     \fi
\ifx \shownote     \undefined \def \shownote      #1{#1}          \fi
\ifx \showarticletitle \undefined \def \showarticletitle #1{#1}   \fi
\ifx \showURL      \undefined \def \showURL       {\relax}        \fi
\providecommand\bibfield[2]{#2}
\providecommand\bibinfo[2]{#2}
\providecommand\natexlab[1]{#1}
\providecommand\showeprint[2][]{arXiv:#2}

\bibitem[\protect\citeauthoryear{Biamonte, Wittek, Pancotti, Rebentrost, Wiebe,
  and Lloyd}{Biamonte et~al\mbox{.}}{2017}]%
        {biamonte2017quantum}
\bibfield{author}{\bibinfo{person}{Jacob Biamonte}, \bibinfo{person}{Peter
  Wittek}, \bibinfo{person}{Nicola Pancotti}, \bibinfo{person}{Patrick
  Rebentrost}, \bibinfo{person}{Nathan Wiebe}, {and} \bibinfo{person}{Seth
  Lloyd}.} \bibinfo{year}{2017}\natexlab{}.
\newblock \showarticletitle{Quantum machine learning}.
\newblock \bibinfo{journal}{\emph{Nature}} \bibinfo{volume}{549},
  \bibinfo{number}{7671} (\bibinfo{year}{2017}).
\newblock


\bibitem[\protect\citeauthoryear{Crooks}{Crooks}{2019}]%
        {crooks2019gradients}
\bibfield{author}{\bibinfo{person}{Gavin~E Crooks}.}
  \bibinfo{year}{2019}\natexlab{}.
\newblock \showarticletitle{Gradients of parameterized quantum gates using the
  parameter-shift rule and gate decomposition}.
\newblock \bibinfo{journal}{\emph{arXiv:1905.13311}} (\bibinfo{year}{2019}).
\newblock


\bibitem[\protect\citeauthoryear{Han, Mao, and Dally}{Han
  et~al\mbox{.}}{2015}]%
        {han2015deep}
\bibfield{author}{\bibinfo{person}{Song Han}, \bibinfo{person}{Huizi Mao},
  {and} \bibinfo{person}{William~J Dally}.} \bibinfo{year}{2015}\natexlab{}.
\newblock \showarticletitle{Deep compression: Compressing deep neural networks
  with pruning, trained quantization and huffman coding}.
\newblock \bibinfo{journal}{\emph{arXiv preprint arXiv:1510.00149}}
  (\bibinfo{year}{2015}).
\newblock


\bibitem[\protect\citeauthoryear{Harrow, Hassidim, and Lloyd}{Harrow
  et~al\mbox{.}}{2009}]%
        {harrow2009quantum}
\bibfield{author}{\bibinfo{person}{Aram~W Harrow}, \bibinfo{person}{Avinatan
  Hassidim}, {and} \bibinfo{person}{Seth Lloyd}.}
  \bibinfo{year}{2009}\natexlab{}.
\newblock \showarticletitle{Quantum algorithm for linear systems of equations}.
\newblock \bibinfo{journal}{\emph{Physical review letters}}
  \bibinfo{volume}{103}, \bibinfo{number}{15} (\bibinfo{year}{2009}),
  \bibinfo{pages}{150502}.
\newblock


\bibitem[\protect\citeauthoryear{Havl{\'\i}{\v{c}}ek
  et~al\mbox{.}}{Havl{\'\i}{\v{c}}ek et~al\mbox{.}}{2019}]%
        {havlivcek2019supervised}
\bibfield{author}{\bibinfo{person}{Vojt{\v{e}}ch Havl{\'\i}{\v{c}}ek}
  {et~al\mbox{.}}} \bibinfo{year}{2019}\natexlab{}.
\newblock \showarticletitle{Supervised learning with quantum-enhanced feature
  spaces}.
\newblock \bibinfo{journal}{\emph{Nature}} \bibinfo{volume}{567},
  \bibinfo{number}{7747} (\bibinfo{year}{2019}), \bibinfo{pages}{209--212}.
\newblock


\bibitem[\protect\citeauthoryear{Hsieh, Wu, Huang, Goan, and Li}{Hsieh
  et~al\mbox{.}}{2020}]%
        {hsieh2020realistic}
\bibfield{author}{\bibinfo{person}{Cheng-Yun Hsieh}, \bibinfo{person}{Chen-Hung
  Wu}, \bibinfo{person}{Chia-Hsien Huang}, \bibinfo{person}{His-Sheng Goan},
  {and} \bibinfo{person}{James Chien~Mo Li}.} \bibinfo{year}{2020}\natexlab{}.
\newblock \showarticletitle{Realistic fault models and fault simulation for
  quantum dot quantum circuits}. In \bibinfo{booktitle}{\emph{2020 57th
  (DAC)}}. IEEE, \bibinfo{pages}{1--6}.
\newblock


\bibitem[\protect\citeauthoryear{IBM}{IBM}{[n.\,d.]}]%
        {ibm_2021}
\bibfield{author}{\bibinfo{person}{Qiskit IBM}.}
  \bibinfo{year}{[n.\,d.]}\natexlab{}.
\newblock
\newblock


\bibitem[\protect\citeauthoryear{{Lecun}, {Bottou}, {Bengio}, and
  {Haffner}}{{Lecun} et~al\mbox{.}}{1998}]%
        {726791}
\bibfield{author}{\bibinfo{person}{Y. {Lecun}}, \bibinfo{person}{L. {Bottou}},
  \bibinfo{person}{Y. {Bengio}}, {and} \bibinfo{person}{P. {Haffner}}.}
  \bibinfo{year}{1998}\natexlab{}.
\newblock \showarticletitle{Gradient-based learning applied to document
  recognition}.
\newblock \bibinfo{journal}{\emph{Proc. IEEE}} \bibinfo{volume}{86},
  \bibinfo{number}{11} (\bibinfo{year}{1998}), \bibinfo{pages}{2278--2324}.
\newblock


\bibitem[\protect\citeauthoryear{Liang, Wang, Yang, Yang, Shi, and Jiang}{Liang
  et~al\mbox{.}}{2021}]%
        {liang2021can}
\bibfield{author}{\bibinfo{person}{Zhiding Liang}, \bibinfo{person}{Zhepeng
  Wang}, \bibinfo{person}{Junhuan Yang}, \bibinfo{person}{Lei Yang},
  \bibinfo{person}{Yiyu Shi}, {and} \bibinfo{person}{Weiwen Jiang}.}
  \bibinfo{year}{2021}\natexlab{}.
\newblock \showarticletitle{Can Noise on Qubits Be Learned in Quantum Neural
  Network? A Case Study on QuantumFlow}. In \bibinfo{booktitle}{\emph{ICCAD}}.
  IEEE, \bibinfo{pages}{1--7}.
\newblock


\bibitem[\protect\citeauthoryear{Lloyd, Garnerone, and Zanardi}{Lloyd
  et~al\mbox{.}}{2016}]%
        {lloyd2016quantum}
\bibfield{author}{\bibinfo{person}{Seth Lloyd}, \bibinfo{person}{Silvano
  Garnerone}, {and} \bibinfo{person}{Paolo Zanardi}.}
  \bibinfo{year}{2016}\natexlab{}.
\newblock \showarticletitle{Quantum algorithms for topological and geometric
  analysis of data}.
\newblock \bibinfo{journal}{\emph{Nature communications}} \bibinfo{volume}{7},
  \bibinfo{number}{1} (\bibinfo{year}{2016}).
\newblock


\bibitem[\protect\citeauthoryear{Lloyd, Mohseni, and Rebentrost}{Lloyd
  et~al\mbox{.}}{2013}]%
        {lloyd2013quantum}
\bibfield{author}{\bibinfo{person}{Seth Lloyd}, \bibinfo{person}{Masoud
  Mohseni}, {and} \bibinfo{person}{Patrick Rebentrost}.}
  \bibinfo{year}{2013}\natexlab{}.
\newblock \showarticletitle{Quantum algorithms for supervised and unsupervised
  machine learning}.
\newblock \bibinfo{journal}{\emph{arXiv:1307.0411}} (\bibinfo{year}{2013}).
\newblock


\bibitem[\protect\citeauthoryear{Lloyd, Schuld, Ijaz, Izaac, and
  Killoran}{Lloyd et~al\mbox{.}}{2020}]%
        {lloyd2020quantum}
\bibfield{author}{\bibinfo{person}{Seth Lloyd}, \bibinfo{person}{Maria Schuld},
  \bibinfo{person}{Aroosa Ijaz}, \bibinfo{person}{Josh Izaac}, {and}
  \bibinfo{person}{Nathan Killoran}.} \bibinfo{year}{2020}\natexlab{}.
\newblock \showarticletitle{Quantum embeddings for machine learning}.
\newblock \bibinfo{journal}{\emph{arXiv:2001.03622}} (\bibinfo{year}{2020}).
\newblock


\bibitem[\protect\citeauthoryear{Magesan, Gambetta, and Emerson}{Magesan
  et~al\mbox{.}}{2012}]%
        {magesan2012characterizing}
\bibfield{author}{\bibinfo{person}{Easwar Magesan}, \bibinfo{person}{Jay~M
  Gambetta}, {and} \bibinfo{person}{Joseph Emerson}.}
  \bibinfo{year}{2012}\natexlab{}.
\newblock \showarticletitle{Characterizing quantum gates via randomized
  benchmarking}.
\newblock \bibinfo{journal}{\emph{Physical Review A}} \bibinfo{volume}{85},
  \bibinfo{number}{4} (\bibinfo{year}{2012}).
\newblock


\bibitem[\protect\citeauthoryear{Mitarai, Negoro, Kitagawa, and Fujii}{Mitarai
  et~al\mbox{.}}{2018}]%
        {mitarai2018quantum}
\bibfield{author}{\bibinfo{person}{Kosuke Mitarai}, \bibinfo{person}{Makoto
  Negoro}, \bibinfo{person}{Masahiro Kitagawa}, {and} \bibinfo{person}{Keisuke
  Fujii}.} \bibinfo{year}{2018}\natexlab{}.
\newblock \showarticletitle{Quantum circuit learning}.
\newblock \bibinfo{journal}{\emph{Physical Review A}} (\bibinfo{year}{2018}).
\newblock


\bibitem[\protect\citeauthoryear{Nguyen-Meidine et~al\mbox{.}}{Nguyen-Meidine
  et~al\mbox{.}}{2020}]%
        {nguyenmeidine2020progressive}
\bibfield{author}{\bibinfo{person}{Le~Thanh Nguyen-Meidine} {et~al\mbox{.}}}
  \bibinfo{year}{2020}\natexlab{}.
\newblock \bibinfo{title}{Progressive Gradient Pruning for Classification,
  Detection and DomainAdaptation}.
\newblock
\newblock
\showeprint[arxiv]{1906.08746}~[cs.LG]


\bibitem[\protect\citeauthoryear{Preskill}{Preskill}{2018}]%
        {preskill2018quantum}
\bibfield{author}{\bibinfo{person}{John Preskill}.}
  \bibinfo{year}{2018}\natexlab{}.
\newblock \showarticletitle{Quantum Computing in the NISQ era and beyond}.
\newblock \bibinfo{journal}{\emph{Quantum}}  \bibinfo{volume}{2}
  (\bibinfo{year}{2018}), \bibinfo{pages}{79}.
\newblock


\bibitem[\protect\citeauthoryear{Wang, Ding, Gu, Lin, Pan, Chong, and Han}{Wang
  et~al\mbox{.}}{2022a}]%
        {wang2021quantumnas}
\bibfield{author}{\bibinfo{person}{Hanrui Wang}, \bibinfo{person}{Yongshan
  Ding}, \bibinfo{person}{Jiaqi Gu}, \bibinfo{person}{Yujun Lin},
  \bibinfo{person}{David~Z Pan}, \bibinfo{person}{Frederic~T Chong}, {and}
  \bibinfo{person}{Song Han}.} \bibinfo{year}{2022}\natexlab{a}.
\newblock \showarticletitle{QuantumNAS: Noise-adaptive search for robust
  quantum circuits}.
\newblock \bibinfo{journal}{\emph{HPCA}} (\bibinfo{year}{2022}).
\newblock


\bibitem[\protect\citeauthoryear{Wang, Gu, Ding, Li, Chong, Pan, and Han}{Wang
  et~al\mbox{.}}{2022b}]%
        {wang2021roqnn}
\bibfield{author}{\bibinfo{person}{Hanrui Wang}, \bibinfo{person}{Jiaqi Gu},
  \bibinfo{person}{Yongshan Ding}, \bibinfo{person}{Zirui Li},
  \bibinfo{person}{Frederic~T Chong}, \bibinfo{person}{David~Z Pan}, {and}
  \bibinfo{person}{Song Han}.} \bibinfo{year}{2022}\natexlab{b}.
\newblock \showarticletitle{QuantumNAT: Quantum Noise-Aware Training with Noise
  Injection, Quantization and Normalization}.
\newblock \bibinfo{journal}{\emph{DAC}} (\bibinfo{year}{2022}).
\newblock


\bibitem[\protect\citeauthoryear{Wang, Zhang, and Han}{Wang
  et~al\mbox{.}}{2021b}]%
        {wang2021spatten}
\bibfield{author}{\bibinfo{person}{Hanrui Wang}, \bibinfo{person}{Zhekai
  Zhang}, {and} \bibinfo{person}{Song Han}.} \bibinfo{year}{2021}\natexlab{b}.
\newblock \showarticletitle{SpAtten: Efficient sparse attention architecture
  with cascade token and head pruning}. In \bibinfo{booktitle}{\emph{HPCA}}.
  IEEE.
\newblock


\bibitem[\protect\citeauthoryear{Wang, Wang, Cai, Lin, Liu, Wang, Lin, and
  Han}{Wang et~al\mbox{.}}{2020}]%
        {wang2020apq}
\bibfield{author}{\bibinfo{person}{Tianzhe Wang}, \bibinfo{person}{Kuan Wang},
  \bibinfo{person}{Han Cai}, \bibinfo{person}{Ji Lin}, \bibinfo{person}{Zhijian
  Liu}, \bibinfo{person}{Hanrui Wang}, \bibinfo{person}{Yujun Lin}, {and}
  \bibinfo{person}{Song Han}.} \bibinfo{year}{2020}\natexlab{}.
\newblock \showarticletitle{Apq: Joint search for network architecture, pruning
  and quantization policy}. In \bibinfo{booktitle}{\emph{CVPR}}.
\newblock


\bibitem[\protect\citeauthoryear{Wang, Liang, Zhou, et~al\mbox{.}}{Wang
  et~al\mbox{.}}{2021a}]%
        {wang2021exploration}
\bibfield{author}{\bibinfo{person}{Zhepeng Wang}, \bibinfo{person}{Zhiding
  Liang}, \bibinfo{person}{Shanglin Zhou}, {et~al\mbox{.}}}
  \bibinfo{year}{2021}\natexlab{a}.
\newblock \showarticletitle{Exploration of Quantum Neural Architecture by
  Mixing Quantum Neuron Designs}. In \bibinfo{booktitle}{\emph{ICCAD}}. IEEE.
\newblock


\bibitem[\protect\citeauthoryear{Xiao, Rasul, and Vollgraf}{Xiao
  et~al\mbox{.}}{2017}]%
        {xiao2017fashion}
\bibfield{author}{\bibinfo{person}{Han Xiao}, \bibinfo{person}{Kashif Rasul},
  {and} \bibinfo{person}{Roland Vollgraf}.} \bibinfo{year}{2017}\natexlab{}.
\newblock \showarticletitle{Fashion-mnist: a novel image dataset for
  benchmarking machine learning algorithms}.
\newblock \bibinfo{journal}{\emph{arXiv:1708.07747}} (\bibinfo{year}{2017}).
\newblock


\bibitem[\protect\citeauthoryear{Zhang, Wang, Han, and Dally}{Zhang
  et~al\mbox{.}}{2020}]%
        {zhang2020sparch}
\bibfield{author}{\bibinfo{person}{Zhekai Zhang}, \bibinfo{person}{Hanrui
  Wang}, \bibinfo{person}{Song Han}, {and} \bibinfo{person}{William~J Dally}.}
  \bibinfo{year}{2020}\natexlab{}.
\newblock \showarticletitle{SpArch: Efficient architecture for sparse matrix
  multiplication}. In \bibinfo{booktitle}{\emph{HPCA}}. IEEE.
\newblock


\end{thebibliography}
}

\end{document}